\documentclass[fleqn,usenatbib]{mnras}

\usepackage{mathptmx}
\usepackage[T1]{fontenc}

\DeclareRobustCommand{\VAN}[3]{#2}
\let\VANthebibliography\thebibliography
\def\thebibliography{\DeclareRobustCommand{\VAN}[3]{##3}\VANthebibliography}

\usepackage{graphicx}	% Including figure files
\usepackage{amsmath}	% Advanced maths commands
\usepackage{amssymb}	% Extra maths symbols
\usepackage{bm}
\newcommand{\isois}{IS$\odot$IS }
\renewcommand{\aa}{\bm{a} }
\newcommand{\bb}{\bm{b} }

\newcommand{\comm}[1]{#1} %commenti
\title[]{Parker Solar Probe Observations of Helical Structures as Boundaries for Energetic Particles}

\author[F. Pecora et al.]{
F. Pecora,$^{1,2}$\thanks{E-mail: fpecora@udel.edu}
S. Servidio,$^{2}$
A. Greco,$^{2}$
W. H. Matthaeus,$^{1}$
D. J. McComas,$^{3}$
J. Giacalone,$^{4}$
C. J. Joyce,$^{3}$
\newauthor
T. Getachew,$^{3}$
C. M. S. Cohen,$^{5}$
R. A. Leske,$^{5}$
M. E. Wiedenbeck,$^{6}$
R. L. McNutt Jr.,$^{7}$
M. E. Hill,$^{7}$
\newauthor
D. G. Mitchell,$^{7}$
E. R. Christian,$^{8}$
E. C. Roelof,$^{7}$
N. A. Schwadron,$^{9}$ and
S. D. Bale$^{10}$
\\
$^{1}$Department of Physics and Astronomy, University of Delaware, Newark, DE 19716, USA\\
$^{2}$Dipartimento di Fisica, Universit\`a della Calabria, I-87036 Cosenza, Italy\\
$^{3}$Department of Astrophysical Sciences, Princeton University, Princeton, NJ 08544, USA\\
$^{4}$Lunar \& Planetary Laboratory, University of Arizona, Tucson, AZ 85721, USA\\
$^{5}$California Institute of Technology, Pasadena, CA 91125, USA\\
$^{6}$Jet Propulsion Laboratory, California Institute of Technology, Pasadena, CA 91109, USA\\
$^{7}$Johns Hopkins University Applied Physics Laboratory, Laurel, MD 20723, USA\\
$^{8}$NASA/Goddard Space Flight Center, Greenbelt, MD 20771, USA\\
$^{9}$University of New Hampshire, Durham, NH 03824, USA\\
$^{10}$Physics Department, University of California at Berkeley, Berkeley, CA 94720, USA
}

\date{Accepted XXX. Received YYY; in original form ZZZ}

\pubyear{}

\begin{document}
\label{firstpage}
\pagerange{\pageref{firstpage}--\pageref{lastpage}}
\maketitle

\begin{abstract}
Energetic particle transport in the interplanetary medium is known to be affected by magnetic structures. It has been demonstrated for solar energetic particles in near-Earth orbit studies, and also for the more energetic cosmic rays. In this paper, we show observational evidence that intensity variations of solar energetic particles can be correlated with the occurrence of helical magnetic flux tubes and their boundaries. The analysis is carried out using data from Parker Solar Probe orbit 5, in the period 2020 May 24 to June 2. We use FIELDS magnetic field data and energetic particle measurements from the Integrated Science Investigation of the Sun (\isois) suite on the Parker Solar Probe. We identify magnetic flux ropes by employing a real-space evaluation of magnetic helicity, and their potential boundaries using the Partial Variance of Increments method. We find that energetic particles are either confined within or localized outside of helical flux tubes, suggesting that the latter act as transport boundaries for particles, consistent with previously developed viewpoints.
%the viewpoint developed by \citet{ruffolo2003trapping} and \citet{tessein2016local}.
\end{abstract}


%%%%%%%%%%%%%%%%%%%%%%%%%%%%%%%%%%%%%%%%%%%%%%%%%%

%%%%%%%%%%%%%%%%% BODY OF PAPER %%%%%%%%%%%%%%%%%%
\section{Introduction}
\label{sec:Intro}

Traditional, simplified models of energetic particle propagation typically assume homogeneous turbulence generated by fields that are wave-like perturbations or random noise. These scenarios lack what would be called intermittency in turbulence theory. Indeed, a variety of indications, both theoretical \citep{ambrosiano1988test, drake2006electron, dalena2012magnetic, pecora2018ion} and observational \citep{mazur2000interplanetary, tessein2013association, tessein2016local, khabarova2017energetic}, build a case that interactions of particles with turbulence is structured and inhomogeneous. Depending on the topology and connectivity of the magnetic field, these interactions may involve temporary trapping \citep{ruffolo2003trapping}, as well as exclusion from certain regions of space \citep{kittinaradorn2009solar}. In some cases, such as solar energetic particle (SEP) ``dropouts'', the influence of the magnetic structure is dramatic \citep{mazur2000interplanetary}; in other cases, it is more subtle, as in cases where detected flux tube boundaries coincide with ``edges'' of 
SEP events \citep{tessein2016local, khabarova2016small}.

With Parker Solar Probe (PSP) \citep{fox2016solar} now reaching distances closer to the Sun than any previous mission, novel opportunities are available for examination of the relationship between magnetic flux structures and energetic particle populations. In particular, energetic particle (EP) measurements from the Integrated Science Investigation of the Sun (IS$\odot$IS) \citep{mccomas2016integrated} along with FIELDS magnetic field measurements \citep{bale2016fields} and SWEAP plasma moments \citep{kasper2016solar}, are enabling characterization of observations of EPs and their transport properties closer to their sources than ever before possible. 

We make use of a novel compact scheme \citep{pecora2020identification} for the detection of helical magnetic flux tubes. The approach makes use of an efficient real-space method for quantifying magnetic helicity \citep{matthaeus1982evaluation}, in conjunction with the partial variance of increments (PVI) \citep{greco2018partial} to detect boundaries within, or at the edges of, magnetic flux tubes.

We find evidence for helical flux ropes acting as transport boundaries for solar energetic particles. Such influence arises from the sudden appearance of SEP enhancements in the vicinity of helical flux tubes accompanied, at their edges, by clusters of enhanced PVI events. This elaborates on previous findings near 1 au \citep{tessein2016local} and indicates that the channeling of SEPs occurs closer to the Sun than has been previously observed.

The paper is organized as follows: in Sec.~\ref{sec:hmpvi} we describe the techniques used to evaluate magnetic helicity and PVI to detect helical flux ropes and their boundaries. In Sec.~\ref{sec:psp} we present the results obtained from the analysis of PSP orbit 5. Opposite but complementary views of energetic particle transport are shown. Finally, in the last Section, we discuss the results.

\section{Local magnetic helicity and PVI methods}
\label{sec:hmpvi}

To correlate EP populations with helical structures and strong PVI events, we make use of the technique described in \citet{pecora2020identification}. The technique exploits the property that flux ropes are, in general, helical. Indeed, flux ropes can be described as structures with approximate cylindrical symmetry, and magnetic field lines that wind about a central axis. From a quantitative perspective, an appropriate measure is the magnetic helicity of the magnetic fluctuations, 
defined as $H_m = \langle \aa \cdot \bb \rangle$ where $\aa$ is the vector potential associated with magnetic field fluctuations $\bb={\bm \nabla}\times{\bm a}$. The averaging operation $\langle \dots \rangle$ is performed over an appropriate volume \citep{woltjer1958theorem, taylor1974relaxation, matthaeus1982measurement}.

Following the definition of \cite{matthaeus1982measurement}, magnetic helicity is related to the off-diagonal part of the magnetic field autocorrelation tensor and can be estimated by single-spacecraft 1D measurements as described in the following. This definition relies only on the assumption of homogeneity, a requirement that is eventually weakened in implementations of local analyses. This method is basically free of any strong symmetry assumptions and does not depend on complex transformations.

The procedure we proposed recently in \citet{pecora2020identification} calculates a {\it local}  value of $H_m$, based on estimates of the  elements of the magnetic field correlation matrix $R_{ij} = \langle b_i(x)b_j(x+l) \rangle$, at a certain point $x$, with increment $l$, and averaging the local correlator of magnetic fluctuations over a region of width $w_0$ centred about $x$. This procedure, implemented entirely in real space, provides a localized evaluation, as do, for example, certain wavelet transforms. The relevant indices $i, j$ for the calculation of helicity are those that refer to directions perpendicular to the relative solar wind-spacecraft motion. For this interval at $\sim$ 0.3 au,  this ``sweeping direction'' is almost purely radial, due to the super-Alfv\'enic flow of the solar wind. In the Radial-Tangential-Normal (RTN) coordinate system, the sweeping of the solar wind is in the R direction, corresponding to increment lags $l$ also in the direction R. Then, the integral to calculate helicity involves the T and N magnetic field components. For the more general case, see \citet{pecora2020identification}. For the present choice of coordinates, $H_m$ is evaluated explicitly as

\begin{equation}    
    H_m(x,\ell) = \int_0^\ell dl~C(x,l) h(l),
\label{eq:Hm}
\end{equation}
where the integral is performed in scale-space and the chosen value of
$\ell$ represents the largest scale that will contribute to $H_m$. In particular, we choose $\ell$ to be a multiple of the correlation length $\lambda_c$ of the specific interval. $h(l) = \frac{1}{2}\left[ 1 + \cos\left(\frac{2\pi l}{w_0}\right) \right]$ is the Hann window used to smooth estimates to zero at the edges of the investigated data interval, thus avoiding spurious effects of boundary fluctuations. $C(x, l)$ is the correlation function defined as
\begin{equation}
C(x, l) = \frac{1}{w_0} \int_{x-\frac{w_0}{2}}^{x+\frac{w_0}{2}} \left[ b_T(\xi)b_N(\xi+l) -b_N(\xi)b_T(\xi+l)\right] d\xi,
    \label{hm1}
    \end{equation}
again, with increments along R. The interval of local integration $w_0$ is arbitrary, but we typically chose it as an order unity multiple of the scale $\ell$, such as $w_0=2\ell$. The above formulas convert directly to the time domain using the Taylor hypothesis. For a detailed derivation of the theory for helicity determination, see \citet{matthaeus1982evaluation} and \cite{matthaeus1982measurement}. In a subsequent paper, we will describe a statistical analysis that is able to separate helicity values attributable to stochastic magnetic field fluctuations, from those which are due to coherent structures.

It is useful to define a normalized version of the magnetic helicity as

\begin{equation}
\tilde{H}_m = \frac{H_m}{\langle \delta b^2 \rangle \lambda_c},
\label{eq:sig}
\end{equation}
where $\langle \delta b^2 \rangle$ is the fluctuation energy computed from a local average taken on an interval of about a correlation length $\lambda_c$.
\comm{
This measure is useful for two purposes: the first is more practical, as this definition does not involve bulk velocity measurements, which are not always available, and Taylor's hypothesis needs not to be used; the second, it has physical relevance as it quantifies magnetic helicity relative to the local energy content, thus permitting smaller helical structures to be readily detected. It is worthwhile to show both $H_m$ and $\tilde{H}_m$ to have a more comprehensive understanding of the helicity-to-energy ratio of the emerging structures (note for example the flux tube from 05-24 14:00 to 17:00 in Fig.~\ref{fig:20200524}. Note that the normalized helicity is a generalization of the Fourier space dimensionless helicity introduced by \citep{matthaeus1982evaluation}.
}
The correlation length $\lambda_c = V_{sw} \tau_c$ is related to the spacecraft correlation time $\tau_c$, and the solar wind speed $V_{sw}$ via the Taylor hypothesis. Computed in this way, the normalized magnetic helicity becomes more sensitive to local conditions. Note the analogy with the normalized magnetic helicity $\sigma_m$ obtained in the Fourier representation of $H_m$ \citep{matthaeus1982measurement}.

Sharp gradients, frequently current sheets, that often reside at external and internal  boundaries of flux ropes are determined using the PVI \citep{greco2008intermittent,greco2018partial}, defined as

\begin{equation}
    \mbox{PVI}(s, \ell) = \frac{ | \Delta {\bm B}(s,\ell) | }{ \sqrt{ \langle | \Delta {\bm B}(s,\ell) |^2 \rangle } },
    \label{pvieq}
\end{equation}
where $\Delta {\bm B}(s,\ell) = {\bm B}(s+\ell) - {\bm B}(s)$ are the magnetic field vector increments evaluated at scale $\ell$ and the averaging operation $\langle \dots \rangle $ is performed over a suitable interval \citep{servidio2011statistical}. The function can be computed spatially in simulations or in magnetic field time series, again by assuming the Taylor hypothesis. The technique has been extensively validated, both in numerical simulations and observations, to be able to identify magnetic field discontinuities, current sheets, and reconnection events \citep{greco2009statistical, osman2014magnetic, greco2018partial, pecora2019single}.

\section{Parker Solar Probe}
\label{sec:psp}

During PSP orbit 5, from 2020 May 24 to June 2, a sequence of several EP events have been measured, at radial distances from 0.45 to 0.2 au, and have been extensively studied \citep{cohen2020parker, chhiber2020magnetic}. We will analyze selected properties of these events using several PSP data products.  We employ magnetic field data from the MAG instrument on the FIELDS suite,  resampled from the original four samples per cycle (4Hz) to a  60-second resolution. PVI is calculated at this scale.  Magnetic helicity is calculated with varying window sizes scaled to multiples of the regional correlation length $\lambda_c$. As the window size $\ell$ increases (decreases), the helicity measurement (Eq.~\ref{eq:Hm}) becomes sensitive to contributions from larger (smaller) scale flux tubes \citep{pecora2020identification}. Particle measurements are obtained from the \isois EPI--Lo and EPI--Hi instruments.

\begin{figure*}
    \centering
    \includegraphics[width=0.9\textwidth]{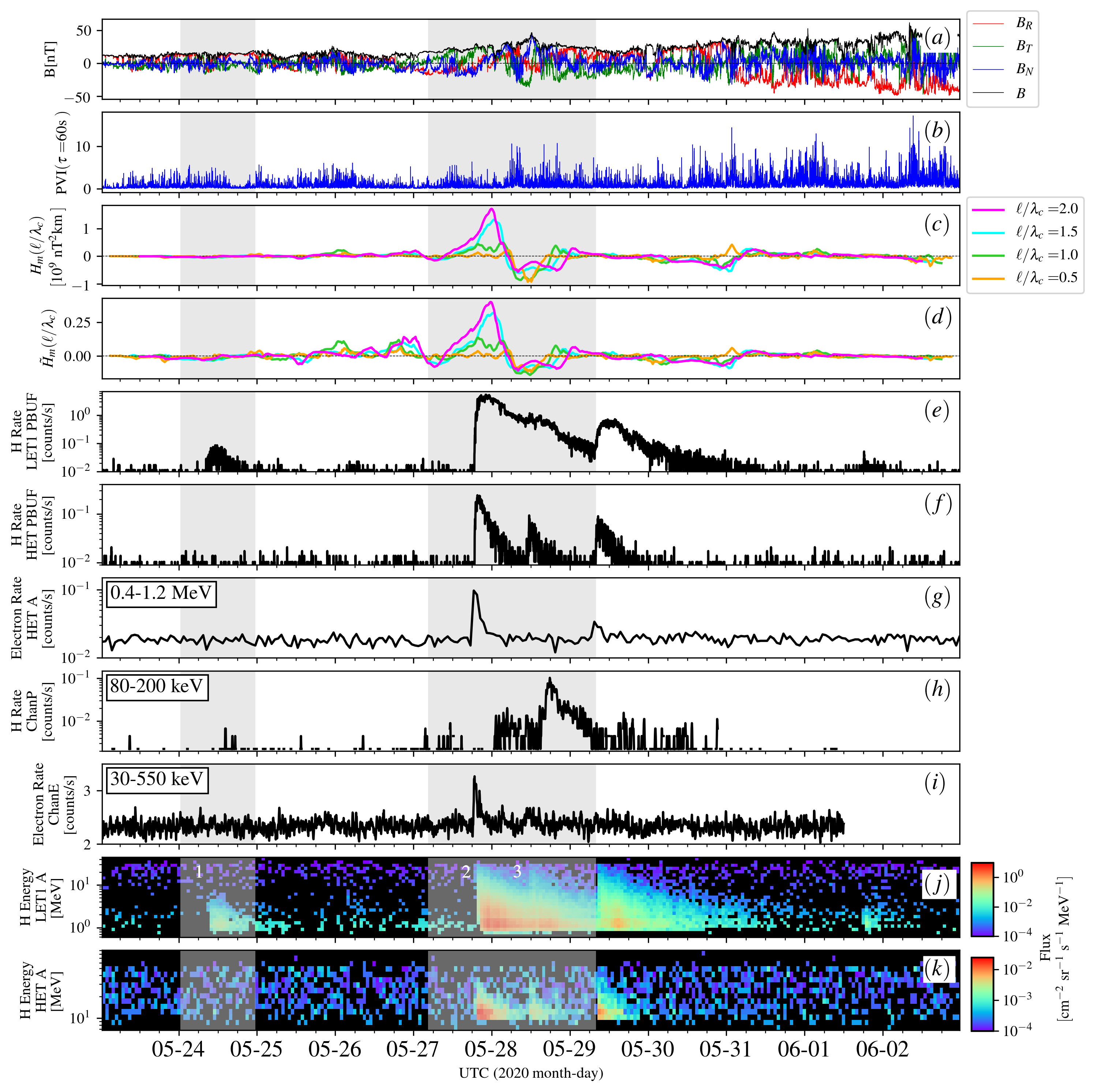}
    \caption{Period from 2020 May 24 to June 2, during which $\lambda_c \sim 6.5 \times 10^6$ km (corresponding to $\tau_c \sim 6$ hours). The stacked panels show (a) magnetic field measured by FIELDS resampled at 60-second cadence, (b) the PVI signal computed with a time lag of 60 seconds, (c) the magnetic helicity of fluctuations at different scales, and (d) its normalized version, (e) proton count rate at 60-second resolution measured by EPI--Hi LET1, (f) proton count rate at 60-second resolution measured by EPI--Hi HET, (g) electron count rate in the energy range 0.4--1.2 MeV at 1-hour resolution measured by EPI--Hi HET A, (h) proton count rate in the energy range 80--200 keV at 300-second resolution measured by EPI--Lo, (i) electron count rate in the energy range 30--550 keV at 300-second resolution measured by EPI--Lo, (j) proton flux at 1-hour resolution measured by EPI--Hi LET1 A, and (k) proton flux at 1-hour resolution measured by EPI--Hi HET A.}
    \label{fig:20200523}
\end{figure*}

\begin{figure*}
    \centering
    \includegraphics[width=0.9\textwidth]{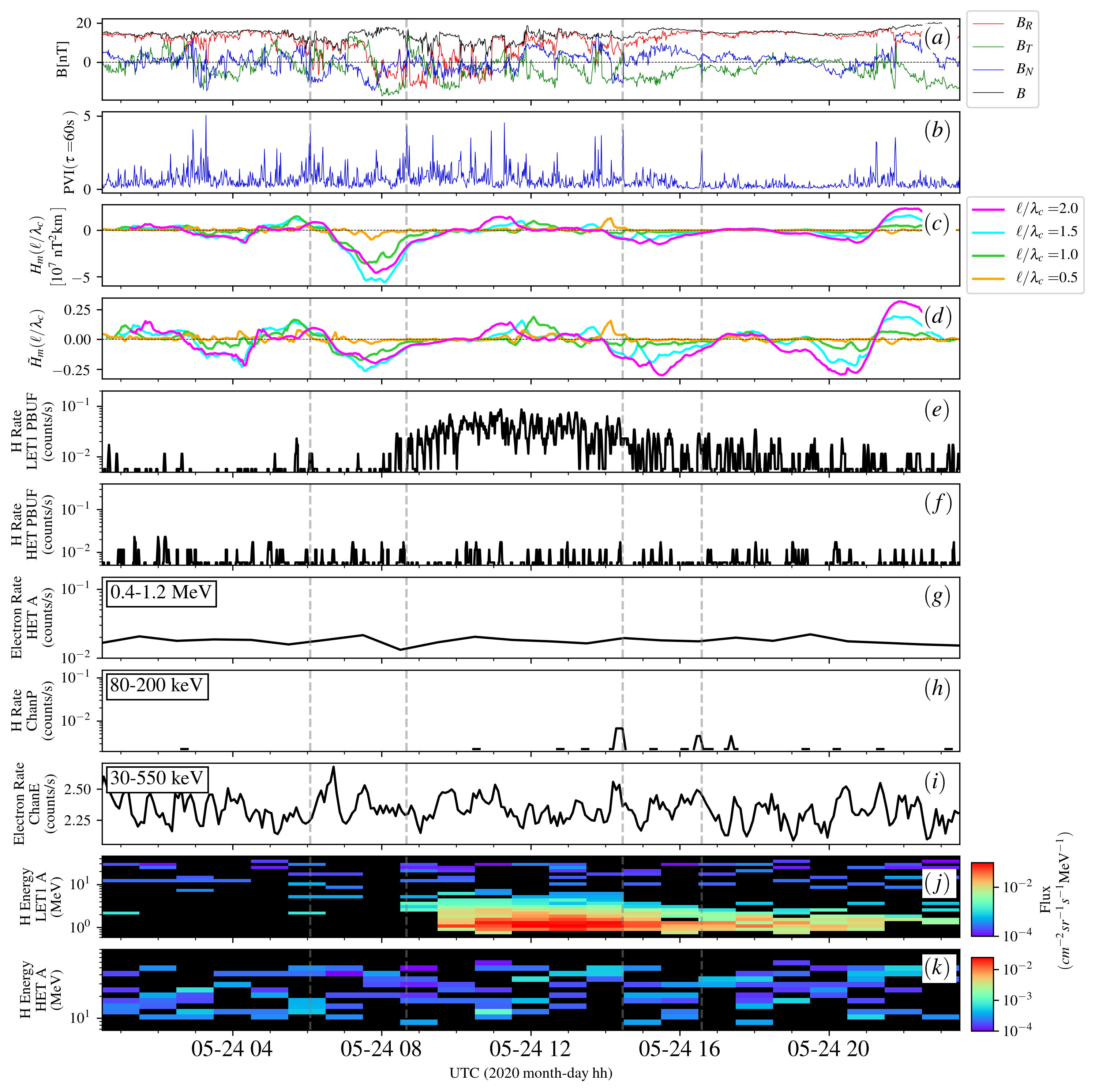}
    \caption{2020 May 24 from 00:30 to 23:30 UTC, $\lambda_c \sim 0.8 \times 10^6$ km ($\tau_c \sim 38$ minutes). The panels are arranged in the same fashion as in Fig.~\ref{fig:20200523}. The energetic proton population, appearing in panel (e), is confined between -- and excluded from -- the leading and trailing negative-helicity peaks. Vertical dashed lines highlight approximate flux tubes boundaries, coinciding with strong PVI events near the edges of helical regions.}
    \label{fig:20200524}
\end{figure*}

\begin{figure*}
    \centering
    \includegraphics[width=0.9\textwidth]{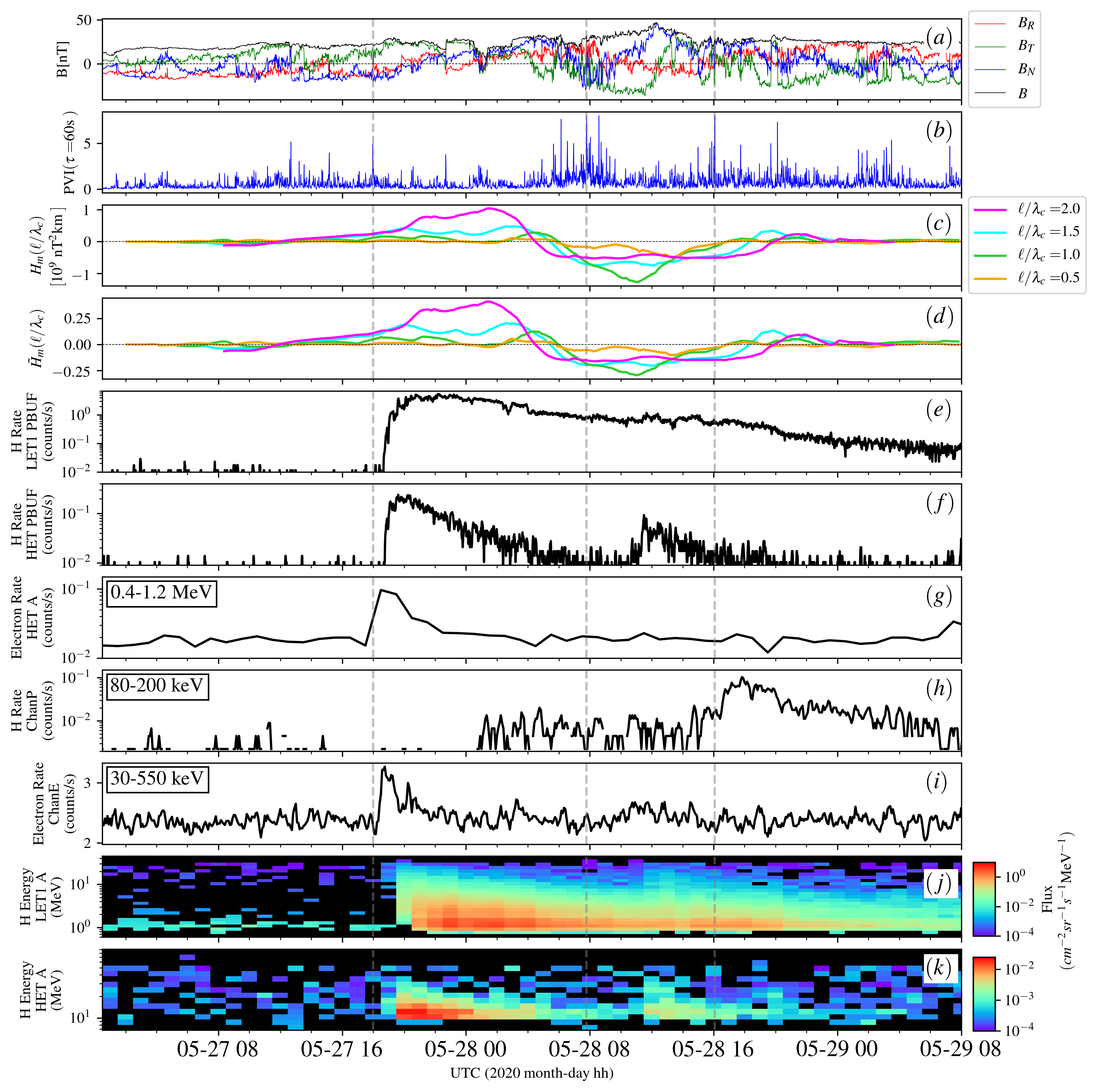}
    \caption{Period from 2020 May 27 04:30 to 29 08:00, during which $\lambda_c \sim 4.0 \times 10^6$ km ($\tau_c \sim 4$ hours). The panels are arranged in the same fashion as in Fig.~\ref{fig:20200523}.
     Both protons and electrons show sudden onsets (panels (e--g), (i--k)) corresponding to the encounter of PSP with the first positive-helicity structure after a strong PVI event. The nature of the magnetic field changes in the period between the two pulses of high-energy particles in panel (f)  -- the sign of helicity changes and there is a burst of PVI activity. This may signal complexity on the transport processes for these apparently distinct events, (see \protect\cite{cohen2020parker}), 
    especially given that the LET1 counts show little change across this transition.
    Vertical dashed lines highlight approximate flux tubes boundaries coinciding with strong PVI events near the edges of helical regions.
    }
    \label{fig:20200527}
\end{figure*}

\comm{
We use count rates from EPI--Lo ChanP (80--200 keV protons) and ChanE (30--550 keV electrons). For EPI--Hi, we use the end A of both the High Energy Telescope (HET A; 0.4--1.2 MeV electrons; 7--60 MeV protons) and the Low Energy Telescope 1 (LET1 A; 1--30 Mev protons).
}
All public data are available from the \isois database and on Coordinated Data Analysis Web (CDAWeb).
\comm{
Priority buffer (PBUF) rates measurements are considered uncalibrated (engineering) data. These are integrated counts measured at different stopping depths within the telescope and cannot be calibrated to fluxes discerning energy ranges. For our use, we summed the ranges R1 through R6, which is similar to integrating over energies. In the following, we do not intend to use these measurements to obtain quantitative estimates, rather they are used to show a better time-resolved envelope of the hourly-averaged fluxes (panels (j) and (k) of Figs.~\ref{fig:20200523}-\ref{fig:20200527}).
}
% Priority buffer (PBUF) rates measurements are not public as they are uncalibrated (engineering) data. These are integrated counts measured at different stopping depths within the telescope and cannot be calibrated to fluxes. For our use, we summed the ranges R1 through R6. In the following, we do not intend to use these measurements to obtain quantitative estimates, rather they are used to show a more resolved envelop of the hourly-averaged fluxes (panels (j) and (k) of the following Figures.)

An overview of the five events occurring during the selected period is shown in Fig.~\ref{fig:20200523}. Even at this scale of 11 days, the sets of measurements show interesting behaviour and a correlation between energetic particles (both protons and electrons) and magnetic field properties. In particular, it is possible to notice that the large helical structure appearing around May 28 encloses both the energetic electrons (panels (g) and (i)) and the higher-energy portion of the energetic proton population (panels (f) and (k)). On the other hand, the LET channel (panel (e)) exhibits fewer structures, suggesting less confinement than its higher energy counterpart HET in panel (f). We will focus on the events labelled as 1, 2, and 3 in the two intervals marked with grey shadings in Fig.~\ref{fig:20200523}, as they give opposite but complementary, views of the ``exclusion'' and ``trapping'' phenomena (see below). We note, in passing, that another event just after the second shaded region, beginning around May 29, 08:00, appears to be nonhelical and possibly nondispersive. That interval will not be discussed in the present paper.  

Figure~\ref{fig:20200524} shows the same quantities as Fig.~\ref{fig:20200523}, but the analysis is performed over the restricted time interval from May 24 00:30 to 23:30 UTC (first shaded region of Fig.~\ref{fig:20200523}) during which $\lambda_c \sim 0.8 \times 10^6$ km (corresponding to $\tau_c \sim 38$ minutes). The restriction to a shorter timescale enhances the smaller-scale helical structures that were obscured before. In this case, the energetic proton population appearing from 8:00 to 15:00 is confined between -- and excluded from -- two helical structures (each indicated with two vertical dashed lines). During this event, the helical field lines appear to act as excluding boundaries for the particles (that may be streaming along ambient solar wind magnetic field lines not organized in a helical fashion), which have suppressed transport across the structures. This kind of exclusionary behaviour is reminiscent of the phenomenon of SEP dropouts that have been associated with topological structures, and that are frequently observed at  1 au and in simulations \citep{mazur2000interplanetary, ruffolo2003trapping, tooprakai2016simulations}.
\comm{
The exclusion of the energetic protons from spatial regions associated with times before 08:00 may be associated with transport effects mediated by the helical tubes that forbid SEPs to access those particular regions of space after their onset.
}
This event appears to be dispersive, as faster particles arrive first (panel (j)), so our suggestion is that the exclusion effect influences which flux tube guided the particle transport from the source region to the point of observation.

The spreading of particles following onset is typically associated with diffusive transport (e.g., \citet{droege2016multi}). In this case, the helical structure near 14:30 may be gradually inhibiting diffusion into the relatively quiet region (in terms of PVI) that resides beyond 16:00. Phenomena such as this have been observed in simulations and have been interpreted as temporary topological trapping \citep{tooprakai2016simulations}, possibly accompanied by suppressed diffusive transport \citep{chuychai2005suppressed}.

The picture given in Fig.~\ref{fig:20200527}, over an approximately two-day period beginning on 2020 May 27, is complementary to the one of Fig.~\ref{fig:20200524}. Energetic particle events 2 and 3, shown here and indicated in panel (j) of Fig.~\ref{fig:20200523}, can be better distinguished in the HET signal of EPI--Hi, in which two separated populations clearly appear (panels (f) and (k)). In LET1, the signal of the first event does not fully decline before the onset of the second one (Fig. \ref{fig:20200527}, panels (e) and (j)). This local analysis shows that there are two adjacent flux tubes of opposite-sign helicity, possibly separated by a strong current sheet individuated by the large-PVI region. One can notice that the dispersive onset of the first EP event, occurring over a period of about an hour, coincides with the appearance of the flux tube, suggesting that the spacecraft has suddenly experienced a different environment, passing from the ambient solar wind to a more confined plasma. In this case, contrary to the previous in Fig.~\ref{fig:20200524}, each energetic population is confined within a different helical structure. 

The lower-energy protons (Fig. \ref{fig:20200527} panel (h)) suggest a more complex history that may intertwine source and transport effects. The activity in panel (h) is not directly associated with the positive-helicity flux rope between 05-27 18:00 and 05-28 08:00 nor with the negative-helicity flux rope between 05-28 08:00 and 05-28 16:00 (the latter being better distinguishable at the scale of one correlation length). Rather, one sees an enhancement of the EPI--Lo 80--200 keV protons in association with the trailing edge of the negative-helicity flux tube. One cannot rule out that the rise in EPI--Lo activity around 05-28 01:00 (panel (h)) may share source region proximity with the EPI--Hi onset near 05-27 18:00 (panels (e) and (f)). Similarly, the larger EPI--Lo increase near 05-28 16:00 (panel (h)) may have an association with the EPI--Hi increase near 05-28 12:00 (panel (f)). If so, the delay in timing would be due to slower propagation of the EPI--Lo particles. However, it is also clear that the flux tubes guiding these lower energy particles to the PSP position have a distinct character, suggesting differences also in transport of the higher and lower energy particles. 

As an aside, we noticed that these energetic particles have gyration radii much smaller than the dimension of the detected flux ropes. The largest gyroradius -- corresponding to a 30 MeV proton in a 25 nT magnetic field -- is about $3 \times 10^4$ km. The average duration of the two flux tubes is about 10 hours, which corresponds to an extension of $\sim 10^7$ km (for an average solar wind speed of 275 km/s on these days).

\comm{
From this comparison of scales, a further consideration arises. Generally, charged particles are expected to follow magnetic field lines; both ions and electrons are subject to electromagnetic field properties. It is clear that the relationship between particle gyroradius and flux tube size can have a significant impact, as can the helicity and energy density of fluctuations. Therefore, each species may behave differently depending on charge sign and energy (rigidity).  As an example, high energy cosmic rays follow the topological properties of the heliosphere \citep{demarco2007numerical}, while lower energy populations are modulated by more local properties \citep{tooprakai2016simulations}. Moreover,  each of these particle properties may influence its behaviour with regard to both transport and energization. The comparison of electron and ion behaviour is a useful step in attempting to unravel and identify these physical effects. 
}

\comm{
In the trapping case of Fig.~\ref{fig:20200527}, even the most energetic ions ($\sim$~30 MeV) seem to be confined within the boundaries of flux tubes. Therefore, we can infer that the same behaviour is expected for lower energy and less massive particles (given that they share source properties as discussed above). The electron signals (panels (g) and (i)) support this view also for opposite-sign particles. Similar conclusions may be drawn for the ``exclusion'' phenomenon of Fig.~\ref{fig:20200524}, even though no high energy ($>$~5MeV) proton or electron measurements are available. In general, we can suggest that both ``exclusion'' and ``trapping'' phenomena are energy-dependent and a comparison of typical scales for both particles and structures need to be considered when describing this type of interaction.
}

\begin{figure}
    \centering
    \includegraphics[width=0.36\textwidth]{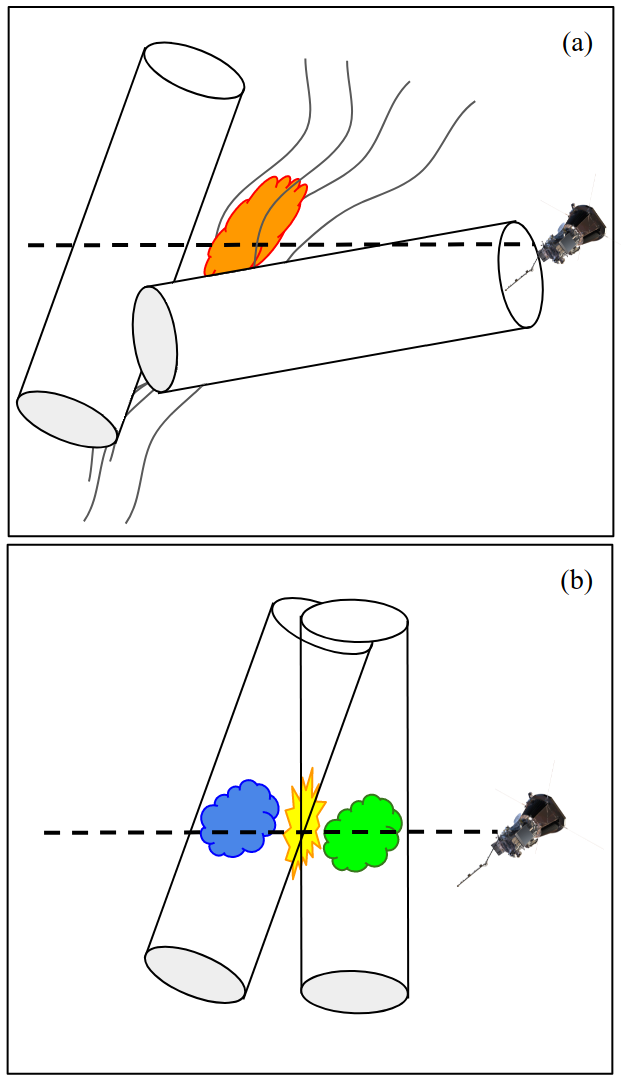}
    \caption{Cartoon illustration of the (a) exclusion and the (b) trapping scenarios. We imagine the flux tubes to be elongated cylinders with magnetic axes aligned with the measured magnetic field. The PSP trajectory, assumed radial, is shown as a dashed line that indicates the sampling of the data. Energetic particles (EPs) are symbolized by clouds. In case (a) EPs transport along file lines outside the two flux tubes. In case (b) the EPs are within the cylindrical tubes, and an interaction region, possibly containing current sheets, is suggested, one that could be responsible for bursting PVI events.}
    \label{fig:cartoon}
\end{figure}

\comm{
A qualitative picture of the exclusion and trapping paradigms is suggested in Fig.~\ref{fig:cartoon}. Several approximations have been made to represent the two scenarios: Flux ropes have their axes directed along the local mean magnetic field, and the approximate path of the PSP trajectory, assumed radial, is shown relative to the flux tube magnetic axes in the two cases.
}

\begin{figure}
    \centering
    \includegraphics[width=0.4\textwidth]{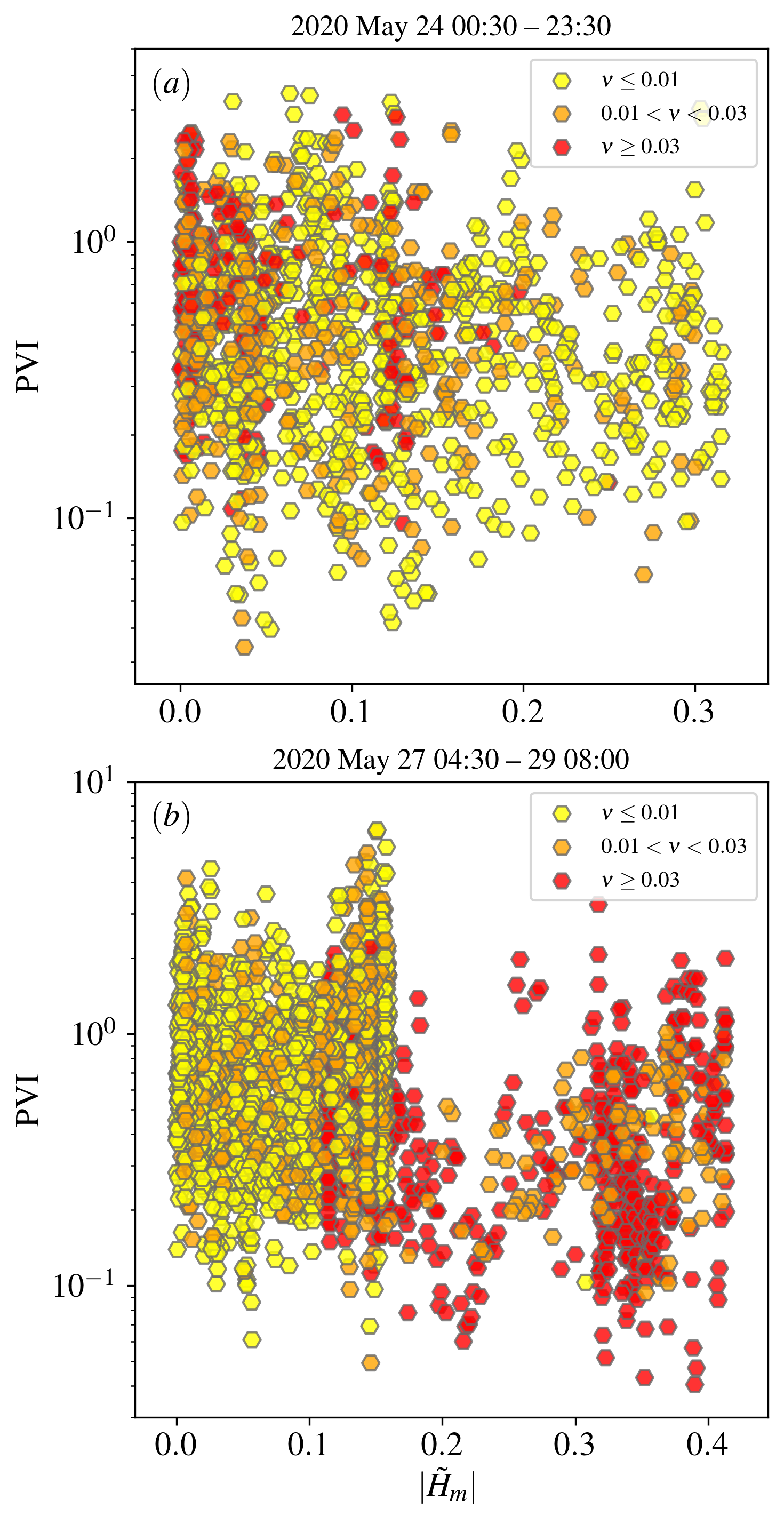}
    \caption{Scatter plot of PVI \textit{vs} $|\tilde{H}_m|$. Larger PVI values are mostly found at smaller $\tilde{H}_m$ (flux tube boundaries). The colour code indicates proton count rates $(\nu)$ of (a) LET1 PBUF (panel (e) of Fig.~\ref{fig:20200524}) and (b) HET PBUF (panel (f) of Fig.~\ref{fig:20200527}.) (a) is illustrative of an exclusion event, while (b) illustrates a trapping event (see text).}
    \label{fig:scatter}
\end{figure}

For a more quantitative visualization of the (anti-)correlation between helical patches and both PVI and EPs in the observed events, we plot PVI versus the absolute value of the normalized helicity (Eq.~\ref{eq:sig}, calculated at the scale of two correlation lengths) as shown in Fig.~\ref{fig:scatter}. The Figure shows data for the May 24 (Fig.~\ref{fig:20200524}) and the May 27-29 intervals (Fig.~\ref{fig:20200527}). It should be apparent from the earlier discussion that the former interval shows particles excluded from a flux tube. For the latter interval,  particles are (at least temporarily) confined or trapped in a flux tube. As discussed before, and as has been confirmed in simulations and observations, high-PVI regions frequently demarcate boundaries of flux ropes \citep{greco2009statistical, pecora2019single}, while relatively large values of $H_m$ characterize the flux tube interiors  \citep{pecora2020identification}. This view is consistent with the panels of Fig.~\ref{fig:scatter}. This alternative and more compact way of representing correlations between PVI and EPs with helical structures will be of even more practical use when longer surveys of several events will be analyzed.

To include EP information, we colour-coded the symbols based on proton count rates, to indicate a general level of energetic particle activity. Specifically, we use LET1 PBUF counts for event 1 of May 24, and HET PBUF counts for events 2 and 3 starting on May 27. The separation in low, mid, and high count rates offers a more compact picture of the scenarios envisioned before. In fact, the distinction between {\it exclusion} and {\it trapping} events is very clear. In the event reported in Fig.~\ref{fig:20200524}, the population of EPs is confined between two consecutive helical structures and excluded from penetrating them; thus, the large count rates in panel (a) of Fig.~\ref{fig:scatter} are confined to lower $\tilde{H}_m$ values. On the contrary, Fig.~\ref{fig:20200527} shows two EP populations confined within two flux ropes and, indeed, in panel (b) of Fig.~\ref{fig:scatter} the large count rates are correlated to large $\tilde{H}_m$ values. Note that the data from the entire respective intervals in Figs.~\ref{fig:20200524} and \ref{fig:20200527} are included in the analysis in Fig.~\ref{fig:scatter}.

\section{Discussion of the Results}

The association of magnetic field properties with energetic particle propagation is a subject of ongoing interest, especially with regard to understanding detailed observations of solar energetic particle events. In particular, there has been renewed interest  \citep{ruffolo2004separation, chuychai2005suppressed, khabarova2014particle, tessein2016local} in understanding how magnetic flux tube structure and topology influence transport as well as acceleration processes \citep{giacalone2000small, drake2006electron}. 

The problem of SEP dropouts has been tackled and addressed in several ways, invoking magnetic field lines connectivity, footpoints motion, and local turbulence properties \citep{mazur1998solar, mazur2000interplanetary, ruffolo2003trapping, chuychai2005suppressed, tooprakai2007temporary, kittinaradorn2009solar, tooprakai2016simulations}. There is also substantial evidence in the literature (e.g., \citet{sanderson1998wind}) that magnetic features associated with the heliospheric current sheet, compression regions, high-speed streams, etc., have a substantial influence on the intensity and occurrence of SEP events. In addition to associations with these large-scale structures, energetic particles (especially electrons) have provided probes of magneto-topology in large magnetic clouds or ICMEs \citep{larson1997using}. The ongoing PSP mission provides the opportunity to examine these issues and the regions of SEP origination closer to the Sun than has been previously possible.

So far, a number of techniques have been used to characterize relevant properties of magnetic field structures, including standard techniques of CME identification \citep{davies2020situ}, wavelet helicity measurements \citep{trenchi2013solar, zhao2020identification}, Grad--Shafranov reconstruction \citep{hu2017grad, pecora2019single}. A recent study \citep{pecora2020identification} combined the use of PVI and a high-performance real-space computation of local magnetic helicity to identify flux tubes in the PSP data, finding results comparable with other methods \citep{zhao2020identification, chen2020small}. Generally speaking,
the present application of these novel statistical techniques extends to smaller-scale magnetic structures than those typically examined in, for example, CME or CIR observations \citep{larson1997using, sanderson1998wind}.

Here, we have applied the $H_m$--PVI technique to selected intervals in which energetic particles are observed by the \isois instruments on PSP. In particular, in the period 2020 May 24 to June 2, during which at least five SEP events are recorded \citep{cohen2020parker}. We focused on two sub-intervals in this period -- one in which energetic particles appear to be confined in the region {\it between} two helical flux ropes, and another in which the particles are confined {\it within} adjacent (and possibly interacting) flux ropes. The combined use of PVI and helicity measurements add details to this characterization; the basic properties of which were previously reported using ACE data and the PVI alone \citep{tessein2013association, tessein2016local, malandraki2019current, khabarova2021current}. 

Both types of events, those described as exclusion events and those described as trapping events, confirm that helical flux structures can provide ``hard walls'' or transport barriers for energetic particles. In fact, numerical experiments have shown in various contexts that both field lines and energetic particles can be temporarily trapped within \citep{chuychai2005suppressed} or temporarily excluded from  \citep{kittinaradorn2009solar} certain regions of space based on their points of origin and the intervening magnetic structures. In a complex environment in which there are many magnetic structures encountered by the particles or field line trajectories, complex patterns can emerge, including the formation of steep gradients. This has been offered as an explanation of dropouts \citep{ruffolo2003trapping, tooprakai2016simulations} as well as solar moss \citep{kittinaradorn2009solar}. 
 
The use of the $H_m$--PVI method has so far produced results that support the interpretations given in the references given above. The PSP observations, here made at $\sim 0.3$ au, are providing a characterization of energetic particles at closer distances to sources in the lower solar atmosphere than has been previously available. The main result of the present work is that within the cluster of SEP events from 2020 May 24 to 2020 May 29, we have been able to find indications of two major types of interactions between energetic particles and magnetic field structures -- namely {\it exclusion} events and {\it trapping} events. These two different, but complementary, empirical descriptions emerging from the events in Fig.~\ref{fig:20200524} and Fig.~\ref{fig:20200527} are not in contrast to one another in terms of basic physical causes; rather they confirm the same vision.
\comm{
The analysis of the two periods suggests that helical flux tubes act as difficult-to-penetrate transport boundaries both for particles that are encapsulated within the structure and are prohibited to leak outside, as well as for those that are found outside and cannot easily gain access to the interior.
}

Particles that are initially outside a strong helical flux tube may have difficulty breaking into the region of helical field lines and populate the core of the flux rope. As in the moss model \citep{kittinaradorn2009solar}, particles can impinge at flux rope boundaries and may get energized by discontinuities and reconnection events. An even more complex scenario observed by PSP is reported recently by \cite{giacalone2021energetic}. In this case, a flux tube-like structure is identified in the vicinity of, and likely passing through, an interplanetary shock and a local sea of energetic particles. The flux tube apparently provides a region of exclusion of particles, perhaps similar to what we have described here, but on a relatively smaller scale. 

The picture proposed here is that EPs are guided by the observed helical flux ropes without influencing the magnetic field. Nonetheless, it is also worth mentioning that in some circumstances EPs can provide non-negligible pressure and actually affect the local topology of the magnetic field. Many classes of MHD equilibria, for example, Grad--Shafranov \citep{hu2017grad} and Chew--Goldberger--Low \citep{chew1956boltzmann}, take into account {\it total} particle pressure, which may include thermal and suprathermal contributions. In cases for which the EP pressure is non-negligible, one might anticipate that EPs confined within a flux tube would cause bulging of the magnetic structure; when EPs are squeezed between helical flux ropes, they might cause inward distortion of the boundaries of the excluding structures. Here the plasma beta based on thermal 
particle
pressure is of order one, so to estimate the potential for an influence of EPs on the magnetic field it suffices to compare EP pressure to magnetic pressure. We carried out a simple estimate based on the formula given by \cite{lario2015energetic} and using the fluences for these events reported by \cite{cohen2020parker}. We estimate that the EP pressure, associated with events 1, 2, and 3 in Figure \ref{fig:20200523}, is at least five to seven orders of magnitude lower than the magnetic pressure. Thus, we can rule out dynamical changes of the magnetic field due to the EP events in these cases.

The overall picture presented here lends some detail to the well-known fact that charged particle transport is controlled by magnetic fields. The events discussed above have been previously analyzed to determine an effective path length \citep{chhiber2020magnetic}. Dispersion analysis suggests a path length of 0.625 au for transport, while the distance from a likely source to PSP along the mean magnetic field is approximately 0.3 au. It has been suggested \citep{chhiber2020clustering} that the extra distance is due to the meandering of the magnetic field lines that guide the particles. It is also clear that field lines are not independent of one another and tend to form structures, sometimes identified as flux tubes or helical flux ropes, often with discontinuities or current structures separating them at boundaries. In the analysis present above, based on Fig.~\ref{fig:20200527}, we see evidence that several distinct SEPs events have produced particle populations that transport to the point of detection at PSP along separate magnetic flux tubes that exhibit distinct properties. \comm{For a more complete discussion of possible sources of the particles, see \citet{cohen2020parker}.}

\comm{Moreover, as reported by \cite{cohen2020parker}, events 2 and 3 have He/H abundance ratios that differ for about two orders of magnitude. This finding may be consistent with the description given above. Indeed, like protons, He ions have orbits much smaller than the size of flux tubes and are likely to be confined in the same regions. Therefore, the impressive difference in abundances might be explained by confinement associated with the local flux tube topology. Leakage and mixing would have been expected for populations residing in adjacent and interacting (as the PVI suggests) structures, resulting in a lower difference in particle composition.}

The different behaviour of higher- and lower-energy protons (panels (e) and (f)) may seem counterintuitive. In numerical experiments \citep{tooprakai2007temporary,tooprakai2016simulations},
it is clear that higher energy particles tend to escape trapping and exclusion barriers more readily. However, in the present observations, the lower-energy population appears to be less confined.
\comm{
The overall effect, in both simulations and observations, involves an interplay between the topological properties of the source region and, later, on the transport properties. Indeed, the initial state leading to an SEP ``trapping'' event might require the source region to be encapsulated within helical magnetic field lines; on the other hand, no constraint is evident for ``exclusion'' events. Later, while flux tubes and EPs propagate, both trapping and exclusion may influence transport by reducing diffusion into and out of flux tubes.
}

%The overall effect, in both simulations and observations, involves an interplay between the topological properties of the source region and, later, on the transport properties.
For the present case, it is reasonable to think of the source region of very high energy particles (panel (f)) to be more localized in space. The generated population is then confined and advected following the coherent magnetic field lines that were encapsulating the source region. On the other hand, the source region of less energetic particles (panel (e)) may be broader in space. Such particles are then transported by a collection of flux tubes, filling entire regions that can include different helicities and even more dispersed/less coherent field lines within that region. This scenario might account for the apparently greater degree of confinement of the higher energy particles in this case.

Also, particles that are originally confined within a flux tube structure may transport more easily along the tube, and may also experience coherent energization processes due to inner reconnection events or due to flux rope topological evolution, such as contraction or expansion \citep{leroux2015energetic, leroux2015kinetic, leroux2018self, du2018plasma}. Whether energization, or perhaps re-energization \citep{khabarova2017energetic}, actually occurs locally in a given event is an intriguing question, but one that lies outside the current scope of this paper. The present results add direct support for the idea that magnetic field topology and structures can influence the transport properties of energetic particles in space plasmas. This itself presents a challenge on the theoretical front since the most widely used transport theories \citep{jokipii1966cosmic,shalchi2004nonlinear}, with rare exception \citep{chuychai2005suppressed}, assume a completely homogeneous plasma medium. We recall that the above interpretation may be visualized, although in cartoon-like form, in Fig.~\ref{fig:cartoon}.

A final point of discussion relates to the issue of how the channelling of EPs by magnetic structures might change in the sub-Alfv\'enic corona at altitudes lower than the Alfv\'en critical point. It is well known that the structures observed by coronagraphs in this region suggest a highly ordered magnetic field, and presumably low plasma beta (e.g. \citet{deforest2018highly}). Near the Alfv\'en critical region, these orderly flux tubes presumably begin to break up and isotropize \citep{deforest2016fading, ruffolo2020shear}. It is then reasonable to suppose that such more orderly flux tubes would act to more effectively confine EPs, though energy-dependent escape is still anticipated \citep{tooprakai2016simulations}. If particle events are detected by PSP below the Alfv\'en critical region, more strict confinement (or exclusion) of EPs by magnetic structures might be directly observed.

The flexibility of the $H_m$--PVI method provides the possibility to conduct this type of analysis on extended data sets at desired scales. In future works, the growing collection of identified EP-containing flux tubes is expected to provide more statistical insight on both the formation and evolution of helical structures and EP transport in space plasmas.

\section*{Acknowledgements}
This research is partially supported by the Parker Solar Probe mission and the \isois project (contract NNN06AA01C) and a subcontract to the University of Delaware from Princeton University (SUB0000165).  Additional support is acknowledged from the  NASA LWS program  (NNX17AB79G) and the HSR program (80NSSC18K1210 \& 80NSSC18K1648). Parker Solar Probe was designed, built, and is now operated by the Johns Hopkins Applied Physics Laboratory as part of NASA’s Living with a Star (LWS) program (contract NNN06AA01C). S. S. has received funding from the European Unions Horizon 2020 research and innovation programme under grant agreement No. 776262 (AIDA, www.aida-space.eu). We thank the \isois team for its support as well as the FIELDS and SWEAP teams for cooperation.

\section*{Data availability}
The \isois data and visualization tools are available to the community at \url{https://spacephysics.princeton.edu/missions-instruments/isois}; data are also available via the NASA Space Physics Data Facility (\url{https://spdf.gsfc.nasa.gov/}).




\bibliographystyle{mnras}
\bibliography{biblio} % if your bibtex file is called example.bib

\begin{thebibliography}{}
\makeatletter
\relax
\def\mn@urlcharsother{\let\do\@makeother \do\$\do\&\do\#\do\^\do\_\do\%\do\~}
\def\mn@doi{\begingroup\mn@urlcharsother \@ifnextchar [ {\mn@doi@}
  {\mn@doi@[]}}
\def\mn@doi@[#1]#2{\def\@tempa{#1}\ifx\@tempa\@empty \href
  {http://dx.doi.org/#2} {doi:#2}\else \href {http://dx.doi.org/#2} {#1}\fi
  \endgroup}
\def\mn@eprint#1#2{\mn@eprint@#1:#2::\@nil}
\def\mn@eprint@arXiv#1{\href {http://arxiv.org/abs/#1} {{\tt arXiv:#1}}}
\def\mn@eprint@dblp#1{\href {http://dblp.uni-trier.de/rec/bibtex/#1.xml}
  {dblp:#1}}
\def\mn@eprint@#1:#2:#3:#4\@nil{\def\@tempa {#1}\def\@tempb {#2}\def\@tempc
  {#3}\ifx \@tempc \@empty \let \@tempc \@tempb \let \@tempb \@tempa \fi \ifx
  \@tempb \@empty \def\@tempb {arXiv}\fi \@ifundefined
  {mn@eprint@\@tempb}{\@tempb:\@tempc}{\expandafter \expandafter \csname
  mn@eprint@\@tempb\endcsname \expandafter{\@tempc}}}

\bibitem[\protect\citeauthoryear{{Ambrosiano}, {Matthaeus}, {Goldstein}  \&
  {Plante}}{{Ambrosiano} et~al.}{1988}]{ambrosiano1988test}
{Ambrosiano} J.,  {Matthaeus} W.~H.,  {Goldstein} M.~L.,   {Plante} D.,  1988,
  \mn@doi [\jgr] {10.1029/JA093iA12p14383}, 93, 14383

\bibitem[\protect\citeauthoryear{Bale et~al.,}{Bale
  et~al.}{2016}]{bale2016fields}
Bale S.~D.,  et~al., 2016, \mn@doi [Space Science Reviews]
  {10.1007/s11214-016-0244-5}, 204, 49

\bibitem[\protect\citeauthoryear{{Chen} et~al.,}{{Chen}
  et~al.}{2020}]{chen2020small}
{Chen} Y.,  et~al., 2020, \mn@doi [\apj] {10.3847/1538-4357/abb820}, 903, 76

\bibitem[\protect\citeauthoryear{{Chew}, {Goldberger}  \& {Low}}{{Chew}
  et~al.}{1956}]{chew1956boltzmann}
{Chew} G.~F.,  {Goldberger} M.~L.,   {Low} F.~E.,  1956, \mn@doi [Proceedings
  of the Royal Society of London Series A] {10.1098/rspa.1956.0116}, 236, 112

\bibitem[\protect\citeauthoryear{{Chhiber, R.} et~al.,}{{Chhiber, R.}
  et~al.}{2021}]{chhiber2020magnetic}
{Chhiber, R.} et~al., 2021, \mn@doi [A\&A] {10.1051/0004-6361/202039816}, 650,
  A26

\bibitem[\protect\citeauthoryear{Chhiber et~al.,}{Chhiber
  et~al.}{2020}]{chhiber2020clustering}
Chhiber R.,  et~al., 2020, \mn@doi [The Astrophysical Journal Supplement
  Series] {10.3847/1538-4365/ab53d2}, 246, 31

\bibitem[\protect\citeauthoryear{{Chuychai}, {Ruffolo}, {Matthaeus}  \&
  {Rowlands}}{{Chuychai} et~al.}{2005}]{chuychai2005suppressed}
{Chuychai} P.,  {Ruffolo} D.,  {Matthaeus} W.~H.,   {Rowlands} G.,  2005,
  \mn@doi [\apjl] {10.1086/498137}, \href
  {https://ui.adsabs.harvard.edu/abs/2005ApJ...633L..49C} {633, L49}

\bibitem[\protect\citeauthoryear{{Cohen, C. M. S.} et~al.,}{{Cohen, C. M. S.}
  et~al.}{2021}]{cohen2020parker}
{Cohen, C. M. S.} et~al., 2021, \mn@doi [A\&A] {10.1051/0004-6361/202039299},
  650, A23

\bibitem[\protect\citeauthoryear{{Dalena}, {Greco}, {Rappazzo}, {Mace}  \&
  {Matthaeus}}{{Dalena} et~al.}{2012}]{dalena2012magnetic}
{Dalena} S.,  {Greco} A.,  {Rappazzo} A.~F.,  {Mace} R.~L.,   {Matthaeus}
  W.~H.,  2012, \mn@doi [\pre] {10.1103/PhysRevE.86.016402}, 86, 016402

\bibitem[\protect\citeauthoryear{Davies et~al.,}{Davies
  et~al.}{2020}]{davies2020situ}
Davies E.,  et~al., 2020, \mn@doi [A\&A] {10.1051/0004-6361/202040113}

\bibitem[\protect\citeauthoryear{{DeForest}, {Matthaeus}, {Viall}  \&
  {Cranmer}}{{DeForest} et~al.}{2016}]{deforest2016fading}
{DeForest} C.~E.,  {Matthaeus} W.~H.,  {Viall} N.~M.,   {Cranmer} S.~R.,  2016,
  \mn@doi [\apj] {10.3847/0004-637X/828/2/66}, 828, 66

\bibitem[\protect\citeauthoryear{{DeForest}, {Howard}, {Velli}, {Viall}  \&
  {Vourlidas}}{{DeForest} et~al.}{2018}]{deforest2018highly}
{DeForest} C.~E.,  {Howard} R.~A.,  {Velli} M.,  {Viall} N.,   {Vourlidas} A.,
  2018, \mn@doi [\apj] {10.3847/1538-4357/aac8e3}, 862, 18

\bibitem[\protect\citeauthoryear{DeMarco, Blasi  \& Stanev}{DeMarco
  et~al.}{2007}]{demarco2007numerical}
DeMarco D.,  Blasi P.,   Stanev T.,  2007, \mn@doi [Journal of Cosmology and
  Astroparticle Physics] {10.1088/1475-7516/2007/06/027}, 2007, 027

\bibitem[\protect\citeauthoryear{{Drake}, {Swisdak}, {Che}  \& {Shay}}{{Drake}
  et~al.}{2006}]{drake2006electron}
{Drake} J.~F.,  {Swisdak} M.,  {Che} H.,   {Shay} M.~A.,  2006, \mn@doi [\nat]
  {10.1038/nature05116}, 443, 553

\bibitem[\protect\citeauthoryear{{Dr{\"o}ge}, {Kartavykh}, {Dresing}  \&
  {Klassen}}{{Dr{\"o}ge} et~al.}{2016}]{droege2016multi}
{Dr{\"o}ge} W.,  {Kartavykh} Y.~Y.,  {Dresing} N.,   {Klassen} A.,  2016,
  \mn@doi [\apj] {10.3847/0004-637X/826/2/134}, 826, 134

\bibitem[\protect\citeauthoryear{Du, Guo, Zank, Li  \& Stanier}{Du
  et~al.}{2018}]{du2018plasma}
Du S.,  Guo F.,  Zank G.~P.,  Li X.,   Stanier A.,  2018, \mn@doi [The
  Astrophysical Journal] {10.3847/1538-4357/aae30e}, 867, 16

\bibitem[\protect\citeauthoryear{Fox et~al.,}{Fox et~al.}{2016}]{fox2016solar}
Fox N.~J.,  et~al., 2016, \mn@doi [Space Science Reviews]
  {10.1007/s11214-015-0211-6}, 204, 7

\bibitem[\protect\citeauthoryear{{Giacalone}, {Jokipii}  \&
  {Mazur}}{{Giacalone} et~al.}{2000}]{giacalone2000small}
{Giacalone} J.,  {Jokipii} J.~R.,   {Mazur} J.~E.,  2000, \mn@doi [\apjl]
  {10.1086/312564}, 532, L75

\bibitem[\protect\citeauthoryear{{Giacalone} et~al.,}{{Giacalone}
  et~al.}{2021}]{giacalone2021energetic}
{Giacalone} J.,  et~al., 2021, in prep.

\bibitem[\protect\citeauthoryear{{Greco}, {Chuychai}, {Matthaeus}, {Servidio}
  \& {Dmitruk}}{{Greco} et~al.}{2008}]{greco2008intermittent}
{Greco} A.,  {Chuychai} P.,  {Matthaeus} W.~H.,  {Servidio} S.,   {Dmitruk} P.,
   2008, \mn@doi [\grl] {10.1029/2008GL035454}, \href
  {http://adsabs.harvard.edu/abs/2008GeoRL..3519111G} {35, L19111}

\bibitem[\protect\citeauthoryear{{Greco}, {Matthaeus}, {Servidio}, {Chuychai}
  \& {Dmitruk}}{{Greco} et~al.}{2009}]{greco2009statistical}
{Greco} A.,  {Matthaeus} W.~H.,  {Servidio} S.,  {Chuychai} P.,   {Dmitruk} P.,
   2009, \mn@doi [The Astrophysical Journall] {10.1088/0004-637X/691/2/L111},
  \href {http://adsabs.harvard.edu/abs/2009ApJ...691L.111G} {691, L111}

\bibitem[\protect\citeauthoryear{{Greco}, {Matthaeus}, {Perri}, {Osman},
  {Servidio}, {Wan}  \& {Dmitruk}}{{Greco} et~al.}{2018}]{greco2018partial}
{Greco} A.,  {Matthaeus} W.~H.,  {Perri} S.,  {Osman} K.~T.,  {Servidio} S.,
  {Wan} M.,   {Dmitruk} P.,  2018, \mn@doi [\ssr] {10.1007/s11214-017-0435-8},
  214, 1

\bibitem[\protect\citeauthoryear{{Hu}}{{Hu}}{2017}]{hu2017grad}
{Hu} Q.,  2017, \mn@doi [Sci.~China Earth Sciences] {doi:
  10.1007/s11430-017-9067-2}, \href
  {http://219.238.6.215/publisher/scp/journal/SCES/doi/10.1007/s11430-017-9067-2?slug=full
  text} {60, 1466}

\bibitem[\protect\citeauthoryear{{Jokipii}}{{Jokipii}}{1966}]{jokipii1966cosmic}
{Jokipii} J.~R.,  1966, \mn@doi [\apj] {10.1086/148912}, 146, 480

\bibitem[\protect\citeauthoryear{Kasper et~al.,}{Kasper
  et~al.}{2016}]{kasper2016solar}
Kasper J.~C.,  et~al., 2016, \mn@doi [Space Science Reviews]
  {10.1007/s11214-015-0206-3}, 204, 131

\bibitem[\protect\citeauthoryear{Khabarova \& Zank}{Khabarova \&
  Zank}{2017}]{khabarova2017energetic}
Khabarova O.~V.,  Zank G.~P.,  2017, \mn@doi [The Astrophysical Journal]
  {10.3847/1538-4357/aa7686}, 843, 4

\bibitem[\protect\citeauthoryear{{Khabarova}, {Zank}, {Li}, {Malandraki}, {le
  Roux}  \& {Webb}}{{Khabarova} et~al.}{2016}]{khabarova2016small}
{Khabarova} O.~V.,  {Zank} G.~P.,  {Li} G.,  {Malandraki} O.~E.,  {le Roux}
  J.~A.,   {Webb} G.~M.,  2016, \mn@doi [\apj] {10.3847/0004-637X/827/2/122},
  827, 122

\bibitem[\protect\citeauthoryear{{Khabarova} et~al.,}{{Khabarova}
  et~al.}{2021}]{khabarova2021current}
{Khabarova} O.,  et~al., 2021, \mn@doi [\ssr] {10.1007/s11214-021-00814-x},
  217, 38

\bibitem[\protect\citeauthoryear{{Kittinaradorn}, {Ruffolo}  \&
  {Matthaeus}}{{Kittinaradorn} et~al.}{2009}]{kittinaradorn2009solar}
{Kittinaradorn} R.,  {Ruffolo} D.,   {Matthaeus} W.~H.,  2009, \mn@doi [\apjl]
  {10.1088/0004-637X/702/2/L138}, 702, L138

\bibitem[\protect\citeauthoryear{{Lario}, {Decker}, {Roelof}  \&
  {Vi{\~n}as}}{{Lario} et~al.}{2015}]{lario2015energetic}
{Lario} D.,  {Decker} R.~B.,  {Roelof} E.~C.,   {Vi{\~n}as} A.~F.,  2015, in
  Journal of Physics Conference Series. p. 012014,
  \mn@doi{10.1088/1742-6596/642/1/012014}

\bibitem[\protect\citeauthoryear{{Larson} et~al.,}{{Larson}
  et~al.}{1997}]{larson1997using}
{Larson} D.~E.,  et~al., 1997, \mn@doi [Advances in Space Research]
  {10.1016/S0273-1177(97)00453-5}, 20, 655

\bibitem[\protect\citeauthoryear{{Malandraki} et~al.,}{{Malandraki}
  et~al.}{2019}]{malandraki2019current}
{Malandraki} O.,  et~al., 2019, \mn@doi [\apj] {10.3847/1538-4357/ab289a},
  \href {https://ui.adsabs.harvard.edu/abs/2019ApJ...881..116M} {881, 116}

\bibitem[\protect\citeauthoryear{Matthaeus \& Goldstein}{Matthaeus \&
  Goldstein}{1982}]{matthaeus1982measurement}
Matthaeus W.~H.,  Goldstein M.~L.,  1982, \mn@doi [\jgr]
  {10.1029/JA087iA08p06011}, 87, 6011

\bibitem[\protect\citeauthoryear{{Matthaeus}, {Goldstein}  \&
  {Smith}}{{Matthaeus} et~al.}{1982}]{matthaeus1982evaluation}
{Matthaeus} W.~H.,  {Goldstein} M.~L.,   {Smith} C.,  1982, \mn@doi [\prl]
  {10.1103/PhysRevLett.48.1256}, 48, 1256

\bibitem[\protect\citeauthoryear{{Mazur}, {Mason}, {Dwyer}  \& {von
  Rosenvinge}}{{Mazur} et~al.}{1998}]{mazur1998solar}
{Mazur} J.~E.,  {Mason} G.~M.,  {Dwyer} J.~R.,   {von Rosenvinge} T.~T.,  1998,
  \mn@doi [\grl] {10.1029/98GL00410}, 25, 2521

\bibitem[\protect\citeauthoryear{{Mazur}, {Mason}, {Dwyer}, {Giacalone},
  {Jokipii}  \& {Stone}}{{Mazur} et~al.}{2000}]{mazur2000interplanetary}
{Mazur} J.~E.,  {Mason} G.~M.,  {Dwyer} J.~R.,  {Giacalone} J.,  {Jokipii}
  J.~R.,   {Stone} E.~C.,  2000, \mn@doi [\apjl] {10.1086/312561}, 532, L79

\bibitem[\protect\citeauthoryear{McComas et~al.,}{McComas
  et~al.}{2016}]{mccomas2016integrated}
McComas D.~J.,  et~al., 2016, \mn@doi [Space Science Reviews]
  {10.1007/s11214-014-0059-1}, 204, 187

\bibitem[\protect\citeauthoryear{{Osman}, {Matthaeus}, {Gosling}, {Greco},
  {Servidio}, {Hnat}, {Chapman}  \& {Phan}}{{Osman}
  et~al.}{2014}]{osman2014magnetic}
{Osman} K.~T.,  {Matthaeus} W.~H.,  {Gosling} J.~T.,  {Greco} A.,  {Servidio}
  S.,  {Hnat} B.,  {Chapman} S.~C.,   {Phan} T.~D.,  2014, \mn@doi [\prl]
  {10.1103/PhysRevLett.112.215002}, 112, 215002

\bibitem[\protect\citeauthoryear{Pecora, Servidio, Greco, Matthaeus,
  D.~Burgess, Carbone  \& Veltri}{Pecora et~al.}{2018}]{pecora2018ion}
Pecora F.,  Servidio S.,  Greco A.,  Matthaeus W.~H.,  D.~Burgess C. T.~H.,
  Carbone V.,   Veltri P.,  2018, \mn@doi [Journal of Plasma Physics]
  {doi.org/10.1017/S0022377818000995}, 84, 725840601

\bibitem[\protect\citeauthoryear{Pecora, Greco, Hu, Servidio, Chasapis  \&
  Matthaeus}{Pecora et~al.}{2019}]{pecora2019single}
Pecora F.,  Greco A.,  Hu Q.,  Servidio S.,  Chasapis A.~G.,   Matthaeus W.~H.,
   2019, \mn@doi [The Astrophysical Journal Letters]
  {doi.org/10.3847/2041-8213/ab32d9}, 881, L11

\bibitem[\protect\citeauthoryear{Pecora, Servidio, Greco  \& Matthaeus}{Pecora
  et~al.}{2020}]{pecora2020identification}
Pecora F.,  Servidio S.,  Greco A.,   Matthaeus W.~H.,  2020, \mn@doi
  [Astronomy \& Astrophysics] {10.1051/0004-6361/202039639}

\bibitem[\protect\citeauthoryear{{Ruffolo}, {Matthaeus}  \&
  {Chuychai}}{{Ruffolo} et~al.}{2003}]{ruffolo2003trapping}
{Ruffolo} D.,  {Matthaeus} W.~H.,   {Chuychai} P.,  2003, \mn@doi [The
  Astrophysical Journall] {10.1086/379847}, \href
  {http://adsabs.harvard.edu/abs/2003ApJ...597L.169R} {597, L169}

\bibitem[\protect\citeauthoryear{{Ruffolo}, {Matthaeus}  \&
  {Chuychai}}{{Ruffolo} et~al.}{2004}]{ruffolo2004separation}
{Ruffolo} D.,  {Matthaeus} W.~H.,   {Chuychai} P.,  2004, \mn@doi [The
  Astrophysical Journal] {10.1086/423412}, 614, 420

\bibitem[\protect\citeauthoryear{Ruffolo et~al.,}{Ruffolo
  et~al.}{2020}]{ruffolo2020shear}
Ruffolo D.,  et~al., 2020, \mn@doi [The Astrophysical Journal]
  {10.3847/1538-4357/abb594}, 902, 94

\bibitem[\protect\citeauthoryear{{Sanderson} et~al.,}{{Sanderson}
  et~al.}{1998}]{sanderson1998wind}
{Sanderson} T.~R.,  et~al., 1998, \mn@doi [\jgr] {10.1029/97JA02884}, 103,
  17235

\bibitem[\protect\citeauthoryear{{Servidio}, {Greco}, {Matthaeus}, {Osman}  \&
  {Dmitruk}}{{Servidio} et~al.}{2011}]{servidio2011statistical}
{Servidio} S.,  {Greco} A.,  {Matthaeus} W.~H.,  {Osman} K.~T.,   {Dmitruk} P.,
   2011, \mn@doi [Journal of Geophysical Research (Space Physics)]
  {10.1029/2011JA016569}, \href
  {http://adsabs.harvard.edu/abs/2011JGRA..116.9102S} {116, A09102}

\bibitem[\protect\citeauthoryear{{Shalchi}, {Bieber}  \& {Matthaeus}}{{Shalchi}
  et~al.}{2004}]{shalchi2004nonlinear}
{Shalchi} A.,  {Bieber} J.~W.,   {Matthaeus} W.~H.,  2004, \mn@doi [\apj]
  {10.1086/424687}, 615, 805

\bibitem[\protect\citeauthoryear{{Taylor}}{{Taylor}}{1974}]{taylor1974relaxation}
{Taylor} J.~B.,  1974, \mn@doi [\prl] {10.1103/PhysRevLett.33.1139}, 33, 1139

\bibitem[\protect\citeauthoryear{{Tessein}, {Matthaeus}, {Wan}, {Osman},
  {Ruffolo}  \& {Giacalone}}{{Tessein} et~al.}{2013}]{tessein2013association}
{Tessein} J.~A.,  {Matthaeus} W.~H.,  {Wan} M.,  {Osman} K.~T.,  {Ruffolo} D.,
   {Giacalone} J.,  2013, \mn@doi [\apjl] {10.1088/2041-8205/776/1/L8}, 776, L8

\bibitem[\protect\citeauthoryear{{Tessein}, {Ruffolo}, {Matthaeus}  \&
  {Wan}}{{Tessein} et~al.}{2016}]{tessein2016local}
{Tessein} J.~A.,  {Ruffolo} D.,  {Matthaeus} W.~H.,   {Wan} M.,  2016, \mn@doi
  [\grl] {10.1002/2016GL068045}, 43, 3620

\bibitem[\protect\citeauthoryear{{Tooprakai}, {Chuychai}, {Minnie}, {Ruffolo},
  {Bieber}  \& {Matthaeus}}{{Tooprakai} et~al.}{2007}]{tooprakai2007temporary}
{Tooprakai} P.,  {Chuychai} P.,  {Minnie} J.,  {Ruffolo} D.,  {Bieber} J.~W.,
  {Matthaeus} W.~H.,  2007, \mn@doi [\grl] {10.1029/2007GL030672}, 34, L17105

\bibitem[\protect\citeauthoryear{{Tooprakai}, {Seripienlert}, {Ruffolo},
  {Chuychai}  \& {Matthaeus}}{{Tooprakai}
  et~al.}{2016}]{tooprakai2016simulations}
{Tooprakai} P.,  {Seripienlert} A.,  {Ruffolo} D.,  {Chuychai} P.,
  {Matthaeus} W.~H.,  2016, \mn@doi [\apj] {10.3847/0004-637X/831/2/195}, 831,
  195

\bibitem[\protect\citeauthoryear{{Trenchi}, {Bruno}, {Telloni}, {D'amicis},
  {Marcucci}, {Zurbuchen}  \& {Weberg}}{{Trenchi}
  et~al.}{2013}]{trenchi2013solar}
{Trenchi} L.,  {Bruno} R.,  {Telloni} D.,  {D'amicis} R.,  {Marcucci} M.~F.,
  {Zurbuchen} T.~H.,   {Weberg} M.,  2013, \mn@doi [\apj]
  {10.1088/0004-637X/770/1/11}, 770, 11

\bibitem[\protect\citeauthoryear{{Woltjer}}{{Woltjer}}{1958}]{woltjer1958theorem}
{Woltjer} L.,  1958, \mn@doi [Proceedings of the National Academy of Science]
  {10.1073/pnas.44.6.489}, 44, 489

\bibitem[\protect\citeauthoryear{{Zank}, {le Roux}, {Webb}, {Dosch}  \&
  {Khabarova}}{{Zank} et~al.}{2014}]{khabarova2014particle}
{Zank} G.~P.,  {le Roux} J.~A.,  {Webb} G.~M.,  {Dosch} A.,   {Khabarova} O.,
  2014, \mn@doi [The Astrophysical Journal] {10.1088/0004-637X/797/1/28}, \href
  {http://adsabs.harvard.edu/abs/2014ApJ...797...28Z} {797, 28}

\bibitem[\protect\citeauthoryear{Zhao et~al.,}{Zhao
  et~al.}{2020}]{zhao2020identification}
Zhao L.-L.,  et~al., 2020, \mn@doi [The Astrophysical Journal Supplement
  Series] {10.3847/1538-4365/ab4ff1}, 246, 26

\bibitem[\protect\citeauthoryear{le Roux, Webb, Zank  \& Khabarova}{le~Roux
  et~al.}{2015a}]{leroux2015energetic}
le Roux J.~A.,  Webb G.~M.,  Zank G.~P.,   Khabarova O.,  2015a, \mn@doi
  [Journal of Physics: Conference Series] {10.1088/1742-6596/642/1/012015},
  642, 012015

\bibitem[\protect\citeauthoryear{{le Roux}, {Zank}, {Webb}  \& {Khabarova}}{{le
  Roux} et~al.}{2015b}]{leroux2015kinetic}
{le Roux} J.~A.,  {Zank} G.~P.,  {Webb} G.~M.,   {Khabarova} O.,  2015b,
  \mn@doi [The Astrophysical Journal] {10.1088/0004-637X/801/2/112}, \href
  {http://adsabs.harvard.edu/abs/2015ApJ...801..112L} {801, 112}

\bibitem[\protect\citeauthoryear{le Roux, Zank  \& Khabarova}{le~Roux
  et~al.}{2018}]{leroux2018self}
le Roux J.~A.,  Zank G.~P.,   Khabarova O.~V.,  2018, \mn@doi [The
  Astrophysical Journal] {10.3847/1538-4357/aad8b3}, 864, 158

\makeatother
\end{thebibliography}

%\begin{thebibliography}{99}
%\end{thebibliography}

%%%%%%%%%%%%%%%%%%%%%%%%%%%%%%%%%%%%%%%%%%%%%%%%%%

%%%%%%%%%%%%%%%%% APPENDICES %%%%%%%%%%%%%%%%%%%%%
% \appendix
% \section{Cartoon representation}
% \label{sec:appx}
% blablabla

%%%%%%%%%%%%%%%%%%%%%%%%%%%%%%%%%%%%%%%%%%%%%%%%%%

% Don't change these lines
\bsp	% typesetting comment
\label{lastpage}
\end{document}